\documentclass[a4paper,11pt]{article}
\usepackage{jcappub}
\usepackage{slashed}
\usepackage[T1]{fontenc}
\usepackage{adjustbox}
\usepackage{multirow}
\usepackage{hhline}
\usepackage{tabularx,booktabs,caption}
\usepackage{float}
\usepackage{placeins}
\usepackage{xcolor}
\usepackage{color}
\usepackage{caption}
\usepackage{verbatim}
\usepackage{soul}
\usepackage[utf8]{inputenc}
\usepackage{natbib}

\usepackage{relsize}
\usepackage{graphics}
\usepackage{epstopdf}
\usepackage{hyperref}
\usepackage{mathrsfs}
\usepackage{ragged2e}
\usepackage{amssymb}
\usepackage{amsthm}
\usepackage{amsmath}

\usepackage{url}
\usepackage{CJKutf8}

\newcommand{\expt}[1]{\left\langle #1 \right\rangle}
\newcommand{\ie}{$i.e.,~$}

\newcommand{\nn}{\nonumber{\nonumber}}
\newcommand{\abs}[1]{\left| #1 \right|}
\newcommand{\beq}{\begin{equation}}
\newcommand{\eeq}{\end{equation}}
\newcommand{\beqn}{\begin{eqnarray}}
\newcommand{\eeqn}{\end{eqnarray}}
\newcommand{\bea}{\begin{eqnarray}}
\newcommand{\eea}{\end{eqnarray}}

\newcommand{\blue}[1]{\textcolor{blue}{#1}}

\newcommand{\chn}[1]{\begin{CJK*}{UTF8}{gbsn}#1\end{CJK*}}

\title{Stochastic Gravitational Wave Background from PBH-ABH Mergers}
\preprint{UCI-HEP-TR-2021-18}

\author[a]{Wenfeng Cui \chn{(崔文峰)},}
\author[a,b]{Fei Huang \chn{(黄飞)},}
\author[a,c,d,e,f,g]{Jing Shu \chn{(舒菁),}}
\author[h]{and Yue Zhao \chn{(赵悦)}} 

\affiliation[a]{CAS Key Laboratory of Theoretical Physics, Institute of Theoretical Physics,\\
Chinese Academy of Sciences, Beijing 100190, China}
\affiliation[b]{Department of Physics and Astronomy, University of California, Irvine, CA 92697 USA}
\affiliation[c]{School of Physical Sciences, University of Chinese Academy of Sciences,\\
Beijing 100049, China}
\affiliation[d]{CAS Center for Excellence in Particle Physics, Beijing 100049, China}
\affiliation[e]{Center for High Energy Physics, Peking University, Beijing 100871, China}
\affiliation[f]{School of Fundamental Physics and Mathematical Sciences,\\
Hangzhou Institute for Advanced Study, UCAS, Hangzhou 310024, China}
\affiliation[g]{International Centre for Theoretical Physics Asia-Pacific, Beijing/Hangzhou, China}
\affiliation[h]
{Department of Physics and Astronomy, University of Utah, Salt Lake City, UT 84112, USA}

\emailAdd{cuiwenfeng@itp.ac.cn}
\emailAdd{huangf4@uci.edu}
\emailAdd{jshu@itp.ac.cn}
\emailAdd{zhaoyue@physics.utah.edu}

\abstract{
The measurement of gravitational waves produced by binary black-hole mergers at the Advanced LIGO has encouraged extensive studies on the stochastic gravitational wave background.
Recent studies have focused on gravitational wave sources made of the same species, such as mergers from binary primordial black holes or those from binary astrophysical black holes.
In this paper, we study a new possibility --- the stochastic gravitational wave background produced by mergers of one primordial black hole and one astrophysical black hole. Such systems are necessarily present if primordial black holes exist.
We study the isotropic gravitational wave background produced through the history of the Universe. We find it is very challenging to detect such a signal. We also demonstrate that it is improper to treat the gravitational waves produced by such binaries in the Milky Way as a directional stochastic background, due to a very low binary formation rate.

}

\DeclareUnicodeCharacter{2212}{-}
\begin{document} 
\captionsetup[figure]{labelfont={bf},labelformat={default},labelsep=period,name={FIG.}}
\maketitle

\section{Introduction}\label{sec:intro}

The first detection of gravitational waves (GW)  ~\cite{Abbott:2016blz}
by the LIGO and Virgo collaborations \cite{aasi2015advanced,acernese2014advanced} in 2015 has opened a new window to study astrophysics and cosmology. 
Since then, many compact binary coalescence events have been observed, including mergers of binary black holes, binary neutron stars and black hole-neutron star binaries \cite{abbott2019gwtc,abbott2020gwtc,LIGOScientific:2021qlt}.
Due to limited sensitivities, those events are located at relatively low redshift, \ie {$z\lesssim 1$}.
Meanwhile, binary mergers can occur at a much earlier time.
Binary mergers of astrophysical black holes (ABHs) can occur shortly after the formation of first stars.
Primordial black holes (PBHs) can be produced due to large density perturbations in the early Universe. The mergers of PBHs could even start deeply within the radiation-dominated epoch and through almost the entire history of the Universe. 

For an individual GW source at a large redshift, the signal is too weak to be detectable. However, the incoherent superposition of a large number of unresolved sources may constitute an observable stochastic gravitational background (SGWB).
The detection or the absence of such a background can therefore teach us about the properties of the GW sources.
For example, the null detection of the SGWB produced by ABH-ABH mergers can be used to constrain various formation scenarios of binary ABHs \cite{TheLIGOScientific:2016wyq}.
Similarly, the search of SGWB produced by PBH-PBH mergers can also be exploited to study the fraction of dark matter in the form of PBHs \cite{Mandic:2016lcn,Wang:2016ana}.
In addition, when combined with the merger-rate history, the SGWB can also be used to distinguish between the population of ABHs from that of PBHs {\cite{Chen:2018rzo,Mukherjee:2021ags,Mukherjee:2021itf,Bavera:2021wmw}.}

Besides binary mergers, many other sources can also produce the SGWB, for example, astrophysical sources like 
supernovae \cite{Marassi:2009ib,Zhu:2010af,Buonanno:2004tp,Sandick:2006sm} 
{and magnetars \cite{Regimbau:2008nj,Marassi:2010wj,Cheng:2015rja,Cheng:2017kmv,Chowdhury:2021vqn}},
and cosmological sources such as cosmic string \cite{LIGOScientific:2021nrg}, inflation  \cite{Grishchuk:1974ny,Starobinsky:1979ty,Boyle:2007zx,LIGOScientific:2014sej} and first-order phase transitions \cite{Romero:2021kby,Huang:2021rrk}. 
For different types of sources, their characteristic frequency as well as the spectral shape can be very different.
Therefore, various species of GW experiments are needed in order to explore interesting physics in different frequency bands.

Typically, ground-based interferometers have sensitivities at relatively high frequency domain.
For example, LIGO, Virgo and KAGRA aim for signals with frequencies between $\sim 10 - 10^3~\rm Hz$.
On the other hand, space-based GW experiments can observe GW at a much lower frequency. For instance, LISA, Taiji and TianQin \cite{Vitale:2014sla,Audley:2017drz,Hu:2017mde,TianQin:2015yph}, have the optimal sensitivities from $10^{-4}~\rm Hz$ to $10^{-1}~\rm Hz$. 
Further lower frequency gravitational waves can be searched by pulsar timing arrays \cite{Smits:2008cf,Hobbs:2009yy}.
Currently, Advanced LIGO and Advanced Virgo have placed an upper limit on the dimensionless GW energy density, $\Omega_{\rm GW}$, for isotropic background at approximately $\mathcal{O}(10^{-9})$ {\cite{TheLIGOScientific:2016dpb,LIGOScientific:2019vic,Abbott:2021xxi,LIGOScientific:2021psn} at $25~\rm Hz$}. 
Such a result constrains certain scenarios of PBH-PBH mergers \cite{Wang:2016ana}, cosmic string network \cite{LIGOScientific:2021nrg} as well as the strong first-order phase transition at a very high scale \cite{Romero:2021kby,Huang:2021rrk}.

It has been demonstrated that a GW experiment can also be used to look for dark matter candidates, in both ultraheavy (see Ref.~\cite{loi:GWPBH} and references therein) and ultralight \cite{Pierce:2018xmy,Guo:2019ker,LIGOScientific:2021odm,Miller:2020vsl,Grote:2019uvn,Vermeulen:2021epa} mass regions. So far, most of the PBH searches using SGWB are focused on PBH-PBH mergers. 
However if PBHs exist, mergers between a PBH and an astrophysical objects naturally arises. 
{For example, a PBH could merge with an ABH in stellar clusters and reproduce the LIGO/Virgo detection rate if the local overdensity of PBHs is large enough \cite{Kritos:2020wcl}.}
In this paper, we study the SGWB produced by mergers of a PBH and an ABH. All mergers in galaxies with different redshifts contribute to the isotropic SGWB.
{Meanwhile}, coalescences that occur in the Milky Way (MW) can also generate a signal with a preferred direction. Both scenarios are considered in this study.

The paper is organized as follows. 
In Sec.~\ref{sec:iso_bg}, we provide the master formula and all the ingredients for estimating the isotropic SGWB. We compare our result with other sources of SGWB as well as the sensitivities of existing and future experiments.
In Sec.~\ref{sec:MW}, we estimate the PBH-ABH formation rate in the MW and study whether the GW produced is proper to be treated as a contribution to the anisotropic SGWB.
Finally, we conclude in Sec.~\ref{sec:conclusion}.

\section{Isotropic Cosmological Background}\label{sec:iso_bg}
\subsection{GW power spectral density}\label{sec:Omega}
The isotropic SGWB is characterized by GW power spectral density $\Omega_{\rm GW}(\nu)$, which is a dimensionless quantity describing the GW energy density per logarithmic frequency interval,
\beqn
\Omega_{\rm GW}(\nu)\equiv\frac{1}{\rho_{c}}\frac{d  \rho_{\rm GW} }{d\ln \nu}=\frac{\nu}{\rho_{c}}\frac{d\rho_{\rm GW}}{d\nu}\,\,\,,\label{eq:GW_abundance_spectrum}
\eeqn
where $\rho_c=3H_0^2/(8\pi G)$ is the critical energy density of the Universe. $H_0$ and $G$ are the Hubble constant and the gravitational constant. {Here,} $\rho_{\rm GW}$ is the energy density of the GW, and $\nu$ is the frequency of the GW observed today. 

{The GW power spectral density consists of GW radiation emitted throughout the entire history of the Universe. For the SGWB generated by ABH-PBH mergers, $\Omega_{\rm GW}$ can be written as an integral over the redshift:}
\beqn
\Omega_{\rm GW}(\nu)&=&\frac{\nu}{\rho_{c}H_0}\int_{0}^{z_{\rm max}} dz~\frac{R_{\rm AP}(z)}{(1+z)^4E(z)} \frac{dE_{\rm GW}}{d\nu_s}(\nu_s) \,\,\,.\label{eq:master_formula_z}
\eeqn
Here, $\frac{dE_{\rm GW}}{d\nu_s}(\nu_s)$ is the GW radiation energy spectrum of the source, $\nu_s$ is the GW frequency at the time of ABH-PBH merger, and it is related to the frequency at observation as $\nu_s= (1+z)\nu$. 
$R_{\rm AP}(z)$ is the ABH-PBH merger rate, \ie the number of mergers per physical volume per cosmological time.  
At last,  $E(z)$ is related to {the Hubble parameter} as $E(z)\equiv H(z)/H_0$.

Therefore, the calculation of $\Omega_{\rm GW}(\nu)$ boils down to the GW radiation energy spectrum for each ABH-PBH merger $\frac{dE_{\rm GW}}{d\nu_s}$ and the merger rate $R_{\rm AP}({z})$. 
We will present details on how they are calculated in the later sections. 
Notice that we impose an upper limit on redshift, $z_{\rm max}$, in the integral. 
This is because {while PBHs can be formed formed at very large redshift deeply within the radiation-dominated epoch,} ABHs can only appear after stars in galaxies have formed. 
Therefore $z_{\rm max}$ refers to the maximal redshift beyond which there is effectively no ABH and consequently no ABH-PBH merger.

\subsection{GW radiation power spectrum}\label{sec:spectrum}
The evolution of a binary merger can be described by three phases:
the \emph{inspiral}, the \emph{merger} and the \emph{ringdown}.
While the GW radiation from the early inspiral and ringdown phases can be approximated analytically by post-Newtonian expansion and perturbation theory, 
modeling the late inspiral and merger requires solving the Einstein equations numerically.
Using the hybrid GW waveform for non-spinning binaries presented in Ref.~\cite{Ajith:2007kx}, the GW energy spectrum can be written as
\begin{equation}
\frac{dE_{\rm GW}}{d\nu_s}=\frac{(G\pi)^{2/3}M_c^{5/3}}{3}\times
	\begin{cases}
	\nu_s^{-1/3}\,\,\,, & \text{if}~\nu_s <\nu_{\rm merg}\,\,\,;\\
	\nu_{\rm merg}^{-1}\nu_s^{2/3}\,\,\,, & \text{if}~\nu_{\rm merg}\leq\nu_s <\nu_{\rm ring}\,\,\,;\\
	\nu_{\rm merg}^{-1}\nu_{\rm ring}^{-4/3}\nu_s^2\left[\left(\displaystyle\frac{\nu_s-\nu_{\rm ring}}{\sigma/2}\right)^2+1\right]^{-2}\,\,\,, & \text{if}~\nu_{\rm ring}\leq\nu_s <\nu_{\rm cut}\,\,\,,
	\end{cases}\label{eq:gw_radiate_power}
\end{equation}
in which $M_c=(m_1 m_2)^{3/5}/(m_1+m_2)^{1/5}$ is the chirp mass of the binary.
The frequencies $\nu_{\rm merg}$ and $\nu_{\rm ring}$ are boundaries that separate the contributions from different regimes --- the inspiral, the merger and the ringdown stages. The parameter
$\sigma$ characterizes the width of the transition from the merger stage to the ringdown stage, and $\nu_{\rm cut}$ is the cutoff of this template.
The frequency dependent behaviors of these parameters are summarized in a vector form $\vec{\alpha}\equiv\{\nu_{\rm merg},~\nu_{\rm ring},~\sigma,~\nu_{\rm cut}\}$ and can then be further parametrized as
\beq
\alpha_j=\frac{a_j \eta^2+b_j \eta + c_j}{\pi}\times\frac{c^3}{MG}\label{eq:alphas}
\eeq
with $\eta=m_1m_2/(m_1+m_2)^2$ as the symmetric mass ratio, and $M$ as the total mass of the binary. The values of $(a_j,b_j,c_j)$ are listed in Table~\ref{tab:radiation_power}.

\begin{table}\centering
\begin{tabular}{c c c c} 
\hline
\hline
Parameter & $a_j$ & $b_j$ & $c_j$ \\ [0.5ex] 
\hline
$\nu_{\rm merg}$ & $2.9740\times 10^{-1}$ & $4.4810\times 10^{-2}$ & $9.5560\times 10^{-2}$\\ 
$\nu_{\rm ring}$ & $5.9411\times 10^{-1}$ & $8.9794\times 10^{-2}$ & $1.9111\times 10^{-1}$ \\
$\sigma$         & $5.0801\times 10^{-1}$ & $7.7515\times 10^{-2}$ & $2.2369\times 10^{-2}$ \\
$\nu_{\rm cut}$  & $8.4845\times 10^{-1}$ & $1.2848\times 10^{-1}$ & $2.7299\times 10^{-1}$ \\
\hline
\hline
\end{tabular}
\caption{Parameters for the GW radiation power spectrum.}\label{tab:radiation_power}
\end{table}

	
\subsection{Merger Rate}\label{sec:rates}


ABHs follow the star distribution in galaxies. Thus the ABH-PBH merger rate can be calculated by integrating the number density of galaxies with the merger rate in each galaxy halo. To be specific, we have
\beq
R_{\rm AP}({z})=\sum_i\int dM_{h}~\frac{dn_h(z,M_h, i)}{dM_h}~R_{\rm AP}^{\rm halo}(z, M_h, i)\,\,\,,\label{eq:rate}
\eeq
where $i$ indicates the type of a galaxy. It includes disk galaxies and elliptical galaxies in this study.
$R_{\rm AP}^{\rm halo}(z,M_h,i)$ is the merger rate of a type-$i$ halo with mass $M_h$ at redshift $z$, 
and $dn_{h}/dM_h$ is the halo mass function with $n_h$ being the physical number density of halos, which can be obtained through numerical simulations.

In order to estimate the merger rate $R_{\rm AP}^{\rm halo}(z,M_h,i)$, the following ingredients are necessary: 
1) the spatial distribution function of PBHs $n_{P}(\vec{r})$; 
2) the spatial distribution function of ABHs $n_{A}(\vec{r})$;
3) ABH-PBH binary formation probability, characterized by the averaged capture cross section $\langle\sigma_{\rm mer}v_{\rm rel}\rangle$. 
With these at hand, the ABH-PBH merger rate per halo can be written as \footnote{In this formula, the merger rate is identified with the binary formation rate. This is reasonable for the binary formation process that we study here in which the delay between the formation of the binary and the subsequent GW emission is negligible compared to cosmological timescales \cite{Bird:2016dcv}. 
{Such binary formation channel assumed here is consistent with the classical isolated single and binary evolution which follows the star formation rate.}}
\beq
R_{\rm AP}^{\rm halo}=\int_{\rm halo} dV  \int dM_A~n_{P}(M_h,z,\vec{r}) \times\frac{dn_{A}}{dM_A}(M_A,M_h,z, \vec{r}) \times \expt{\sigma_{\rm mer}(M_A,M_P,v_{\rm rel})~v_{\rm rel}} \,,\label{eq:rate_per_halo}
\eeq
where we keep the explicit dependence on the redshift $z$, the halo mass $M_h$, the ABH/PBH masses, as well as the spatial location of the black holes $\vec{r}$ inside the halo.
In the rest of this section, we provide details for these three ingredients and then combine them with the halo mass function to estimate the integrated merger rate.

\subsubsection{PBH distribution}


The mass of the PBH is not well predicted theoretically. 
In this study, we assume it takes a single value,
and we consider two benchmarks, $M_P=1$ and $30~M_{\odot}$. 
Since the PBHs are produced at the very early time of the Universe, its spatial distribution follows that of dark matter.
Assuming PBHs constitute a fraction, $f$, of the total dark matter abundance, the PBH number density can be written as
\beq
n_{P}=\frac{\rho_{P}}{M_P}=f\times\frac{\rho_{\rm dm}}{M_P}\,\,\,.\label{eq:nP}
\eeq
Here we assume the dark-matter distribution $\rho_{\rm dm}$ is described by the NFW profile \cite{Navarro:1996gj},
\beq
\rho_{\rm dm}(r)=\frac{\rho_{0}}{r/R_{s}(1+r/R_s)^2}\,\,\,,\label{eq:NFW}
\eeq
where $r$ is the distance from the center of the halo, $\rho_0$ is the normalization factor and $R_s$ is related to the virial radius via the concentration parameter $C(M_h, z)=R_{\rm vir}/R_s$. 
In this study, we determine the concentration parameter using the fitting formula in Ref.~\cite{Prada:2011jf}
and only consider $C(M_h, z)$ between $0.5$ and $1000$ in order to avoid a divergent result \cite{Mandic:2016lcn}.

The virial radius of a halo is defined through the averaged density within the region. More explicitly, for a halo at redshift $z$, one has
\beq
M_h\simeq\Delta \times \left[\Omega_m (1+z)^3+\Omega_\Lambda \right]\rho_c\times\frac{4\pi}{3}R_{\rm vir}^3\,\,\,, \label{eq:virial_r}
\eeq
where $\Delta$ is typically taken as 200 \cite{White:2000jv}.

\subsubsection{ABH distribution}

The ABH mass ranges from $\sim 5~M_{\odot}$ to a few tens of solar masses \cite{Bailyn:1997xt,Farr:2010tu}.
In order to calculate the distribution of ABHs with respect to different masses, we use the initial mass function (IMF) \cite{Kroupa:2000iv} which describes the number distribution of stars:
\beq
\frac{dN_M}{dM}\propto
\begin{cases}
 M^{-0.3}\,, & 0.01~M_{\odot}\leq M<0.08~M_{\odot}\,\,\,\\
 M^{-1.3}\,, & 0.08~M_{\odot}\leq M<0.5 ~M_{\odot}\,\,\,\\
 M^{-2.3}\,, & 0.5~ M_{\odot}\leq M<100 ~M_{\odot}\,\,\,
\end{cases}.
\eeq
{To determine the fraction of stars that eventually forms black holes, we make a simple assumption that only stars with masses larger than 25 $M_\odot$ become black holes.\footnote{{This assumption follows from the conclusion in Ref.~\cite{Heger:2002by} that a star with metallicity between metal-free and solar metallicity becomes a black hole by supernova mass fallback or direct core collapse if its mass is larger than 25 $M_\odot$, and those with lower masses can only form white dwarfs or neutron stars. However, recent studies show that the relation between neutron star and black-hole formation is more sophisticated and there is not a single critical mass above which black holes can form \cite{OConnor:2010moj,Ugliano:2012fvp,Pejcha:2014wda,Sukhbold:2015wba,Ertl:2019zks,Patton:2020tiy,daSilvaSchneider:2020ddu}.}}}
Moreover, although the formation of a black hole takes a finite amount of time, the lifetime of a star with its mass larger than 25 $M_\odot$ is quite small compared with the cosmological time that we consider in this paper.
Therefore, we ignore the ABH formation time and estimate its number density per solar mass of stellar objects as
\beqn
\frac{\chi_{A}(M_A)}{M_{\odot}} \simeq\frac{\displaystyle\frac{dN_M}{dM}\bigg|_{M=M_A^{\rm pro}}}{\displaystyle\int_{0.08~M_{\odot}}^{100~M_{\odot}}~M\frac{dN_M}{dM}~dM}\,\,\,.\label{eq:ABH_per_solar}
\eeqn
{For simplicity, we follow the similar approach in \cite{Kritos:2020fjw} and assume the remnant ABH only retains $\sim 1/3$ of the initial stellar mass. Therefore, $M_A^{\rm pro}\simeq 3M_A$ stands for the mass of the progenitor associated with an ABH of mass $M_A$.
In reality, the remnant mass can be affected by stellar winds which depend on metallicity and thus also depend on redshift \cite{Vink:2001cg,Graefener:2008pq,Vink:2011kd,2015MNRAS.452.1068C}.}
Integrating the equation above, we find that the averaged number of ABHs per solar mass is $\sim~2\times10^{-3}$.
For a galaxy with mass distribution $\rho_G$, we thus have the spatial distribution of the ABH number density as a function of the ABH mass as
\beq
\frac{dn_A}{dM_A}=\frac{\chi_{A}(M_A)}{M_{\odot}}\times\rho_{G}\,\,\,.
\eeq
{Notice that this relation assumes that all stars and thus all ABHs are formed in isolation and in the field. 
In practice, the majority of them are born in binary systems \cite{Sana:2012px,2017ApJS..230...15M}, which might further affect our assumption of two-body capture. We shall reserve the consideration of such effect for future work.}

The distribution $\rho_{G}$ depends on multiple aspects of a galaxy, \ie the redshift, the halo mass and also the type of the galaxy.
For elliptical galaxies, we take the Hernquist Model \cite{Hernquist:1990be}
\beq
\rho_G(r)=\frac{C_E}{2\pi} \frac{R_c}{r(r+R_c)^3}\,\,\,,\label{eq:elliptical}
\eeq
in which $r$ is the radius in spherical coordinates, and the core radius $R_c$ can be determined by its relation to the half-light (half-mass) radius $(\sqrt{2}+1)R_c=R_{1/2}$.
For disk galaxies, the mass distribution is approximately described by a double exponential form \cite{2010gfe..book.....M}:
\beq
\rho_G(R, h)=C_D \exp(-R/R_D)\exp(-\abs{h}/h_D)\,\,\,,\label{eq:disk}
\eeq
where $R$ and $h$ are the radius and height in cylindrical coordinates. $R_D$ and $h_D$ are related to the half-light radius of the halo hosting the galaxy as $1.68 R_D\simeq R_{1/2}$, 
 and $h_D\simeq R_D/10$.

In both cases, we have $R_{1/2}=\lambda R_{\rm vir}$ with $\lambda\simeq 0.015$ \cite{Kravtsov:2012jn}.
Moreover, the normalization factor $C_{D,E}$ in both profiles are determined by 
\beq
\int_{\rm halo} dV \rho_G=M_s\,\,\,,
\eeq
where $M_s$ is the total stellar mass in the galaxy.

\begin{figure}
    \centering
    \includegraphics[width=0.7\textwidth]{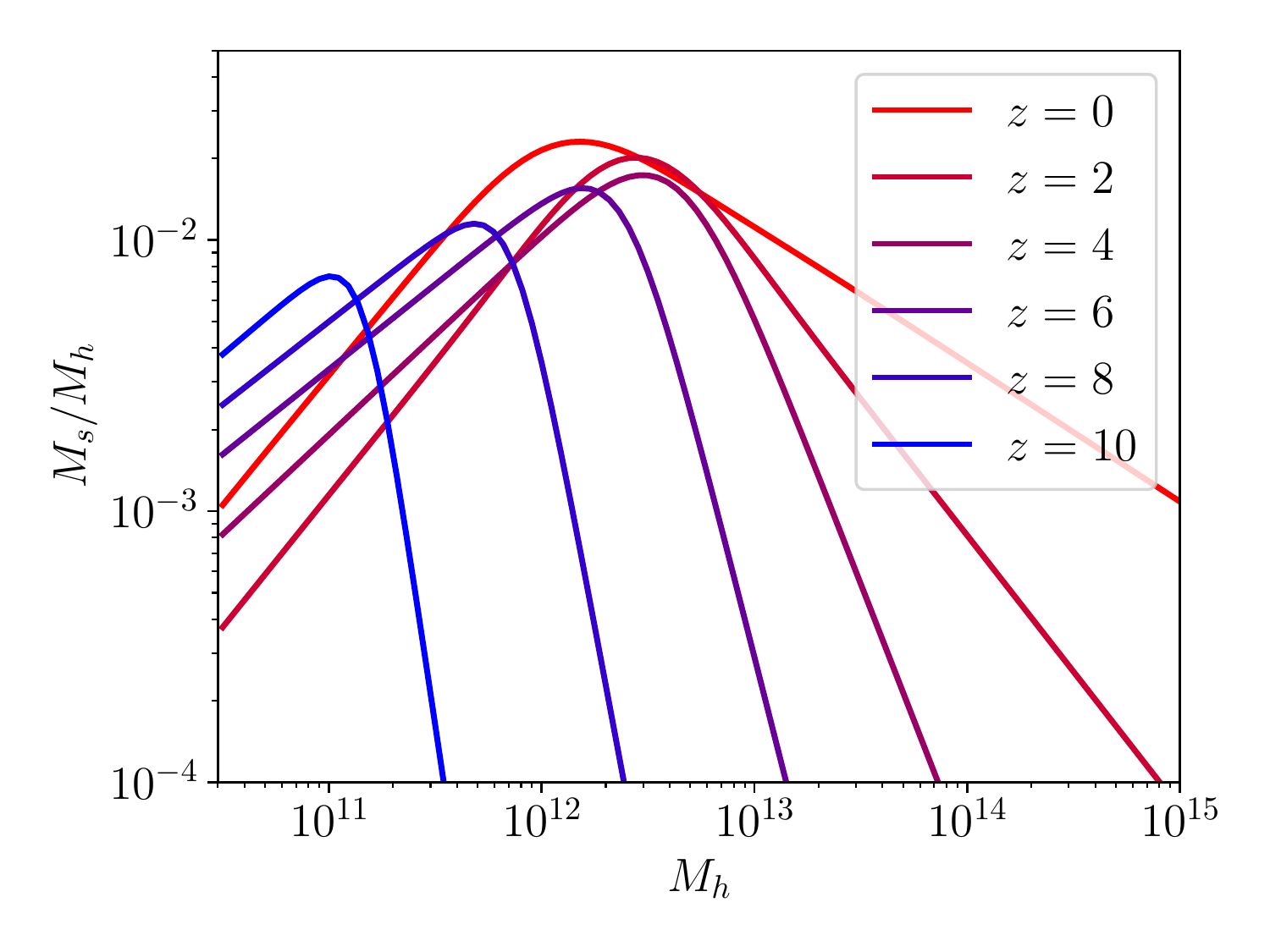}
    \caption{Stellar-halo mass relation.}
    \label{fig:Ms_Mh}
\end{figure}

We determine the total stellar mass using the stellar-halo mass relation provided in Ref.~\cite{Behroozi:2019kql}.
In particular, the stellar-halo mass relations are parametrized as 
\beq
\log_{10} \left(\frac{M_s}{M_1}\right)=\epsilon - \log_{10}\left( 10^{-\alpha x} +10^{-\beta x} \right) +\gamma \exp\left[-\frac{1}{2}\left(\frac{x}{\delta} \right)^2 \right]\,,
\eeq
in which $x=\log_{10}(M_h/M_1)$.
The parameters scales with redshift as
\beqn
\log_{10}\left(\frac{M_1}{M_{\odot}} \right)&=&M_0+M_a(a-1)-M_{\ln a}\ln a + M_z z \,,\\
\epsilon&=&\epsilon_0+\epsilon_a(a-1)-\epsilon_{\ln a}\ln a + \epsilon_z z\,,\\
\alpha&=&\alpha_0+\alpha_a(a-1)-\alpha_{\ln a}\ln a + \alpha_z z\,,\\
\beta&=&\beta_0+\beta_a(a-1) + \beta_z z\,,\\
\log_{10}\gamma&=&\gamma_0+\gamma_a(a-1) + \gamma_z z\,,\\
\delta&=&\delta_0\,,
\eeqn
in which $a\equiv 1/(1+z)$ is the scale factor, and the values of the additional parameters are summarized in Table~\ref{tab:Ms_Mh}. 
A few examples of the stellar-halo mass relation at different redshifts are shown in FIG.~\ref{fig:Ms_Mh}.

\begin{table}\centering
 \begin{tabular}{c c c c c c c c} 
 \hline\hline
 $M_0$ & $M_a$ & $M_{\ln a}$ & $M_z$ & $\epsilon_0$ & $\epsilon_a$ & $\epsilon_{\ln a}$ & $\epsilon_z$ \\ 
 $12.06$ & $4.609$ & $4.525$ & $-0.756$ &
 $-1.459$ & $1.515$ & $1.249$ & $-0.214$\\ 
 \hline
 $\alpha_0$ & $\alpha_a$ & $\alpha_{\ln a}$ & $\alpha_z$ &
 $\beta_0$ & $\beta_a$ & - & $\beta_z$ \\
 $1.972$ & $-2.523$ & $-1.868$ & $0.188$&
 $0.488$ & $-0.965$ & - & $-0.569$ \\
 \hline
 $\gamma_0$ & $\gamma_a$ & - & $\gamma_z$ &
 $\delta_0$ & - & - & -\\
 $-0.958$ & $-2.230$ & - & $-0.706$&
 $0.391$ & - & - & -\\
 \hline\hline
\end{tabular}
\caption{Parameters for the stellar-halo mass relation.}\label{tab:Ms_Mh}
\end{table}

In our analysis, we consider two limits where all galaxies in the Universe are either disk galaxies or elliptical galaxies. The reality should lie between the results from these two limits.

\subsubsection{ABH-PBH binary formation probability}

We estimate the ABH-PBH binary merger rate through gravitational capture process. 
As an ABH and a PBH pass each other, the gravitational wave radiation takes away some amount of the energy of the system. 
If the energy loss is large enough and brings the total energy of the system below zero, a bound state forms, and the merger will happen soon after. 
For binaries formed via this mechanism,
the characteristic delay time, describing the duration between the binary formation and the final coalescence, depends on the velocity dispersion of the hosting halo. Is typically much shorter (hours for $M_h\sim 10^{12}~M_\odot$ and kyrs for $M_h\sim 10^{6}~M_\odot$ \cite{Bird:2016dcv}) compared to the Hubble time. Therefore we can safely treat it as instantaneous on cosmological time scales.
The gravitational capture cross-section can be estimated as \cite{Mouri:2002mc}
\beq
	\sigma_{\rm mer}(m_i,m_j)=2\pi \left(\frac{85\pi}{6\sqrt{2}}\right)^{2/7}\frac{G^2(m_i+m_j)^{10/7}m_i^{2/7}m_j^{2/7}}{c^{10/7}v_{\rm rel}^{18/7}}\,\,\,,\label{eq:xsection}
\eeq
where $m_i$ and $m_j$ are the masses of two black holes which will be identified as the ABH mass $M_{A}$ and the PBH mass $M_{P}$ respectively, 
and $v_{\rm rel}$ is the relative velocity between these two black holes.


We assume both the ABH and the PBH velocities follow the same Maxwell-Boltzmann distribution with a cutoff at the virial velocity, $v_{\rm vir}\equiv\sqrt{2GM_h/R_{\rm vir}}$, of a halo with mass $M_h$ \cite{Bird:2016dcv,Mandic:2016lcn}:
\beqn
P(v,v_{\rm m})&=&F_0\left[\exp\left(-\frac{v^2}{v_{\rm m}^2}\right) - \exp\left(-\frac{v_{\rm vir}^2}{v_{\rm m}^2}\right) \right]\,\,\,,\label{eq:v_prob}\\
v_{\rm m}&=&\frac{v_{\rm vir}}{\sqrt{2}}\sqrt{\frac{C}{C_m}\frac{g(C_m)}{g(C)}}\,\,\,,\\
g(X)&=&\ln(1+X)-\frac{1}{1+X}\,\,\,,
\eeqn
where $v_{\rm m}$ is the maximum circular velocity in an NFW halo which occurs at $R_m=C_m R_s$ with $C_m=2.1626$, and $F_0$ is the normalization factor so that $\int_0^{v_{\rm vir}} dv~4\pi v^2 P(v,v_{\rm m})=1$.
Therefore, averaged cross-section in a halo is defined as 
\beq
\expt{\sigma_{\rm mer} v}\equiv\int d^3v_1 d^3v_2~ \sigma_{\rm mer} v_{\rm rel} P(v_1,v_m) P(v_2,v_m)\,\,\,,\label{eq:average_xsection}
\eeq
in which $v_{1}$ and $v_2$ are the velocities of ABHs and PBHs respectively, and $v_{\rm rel}=\abs{\vec{v}_1-\vec{v}_2}$.

\subsubsection{Halo mass function}

\begin{figure}
    \centering
    \includegraphics[width=0.7\textwidth]{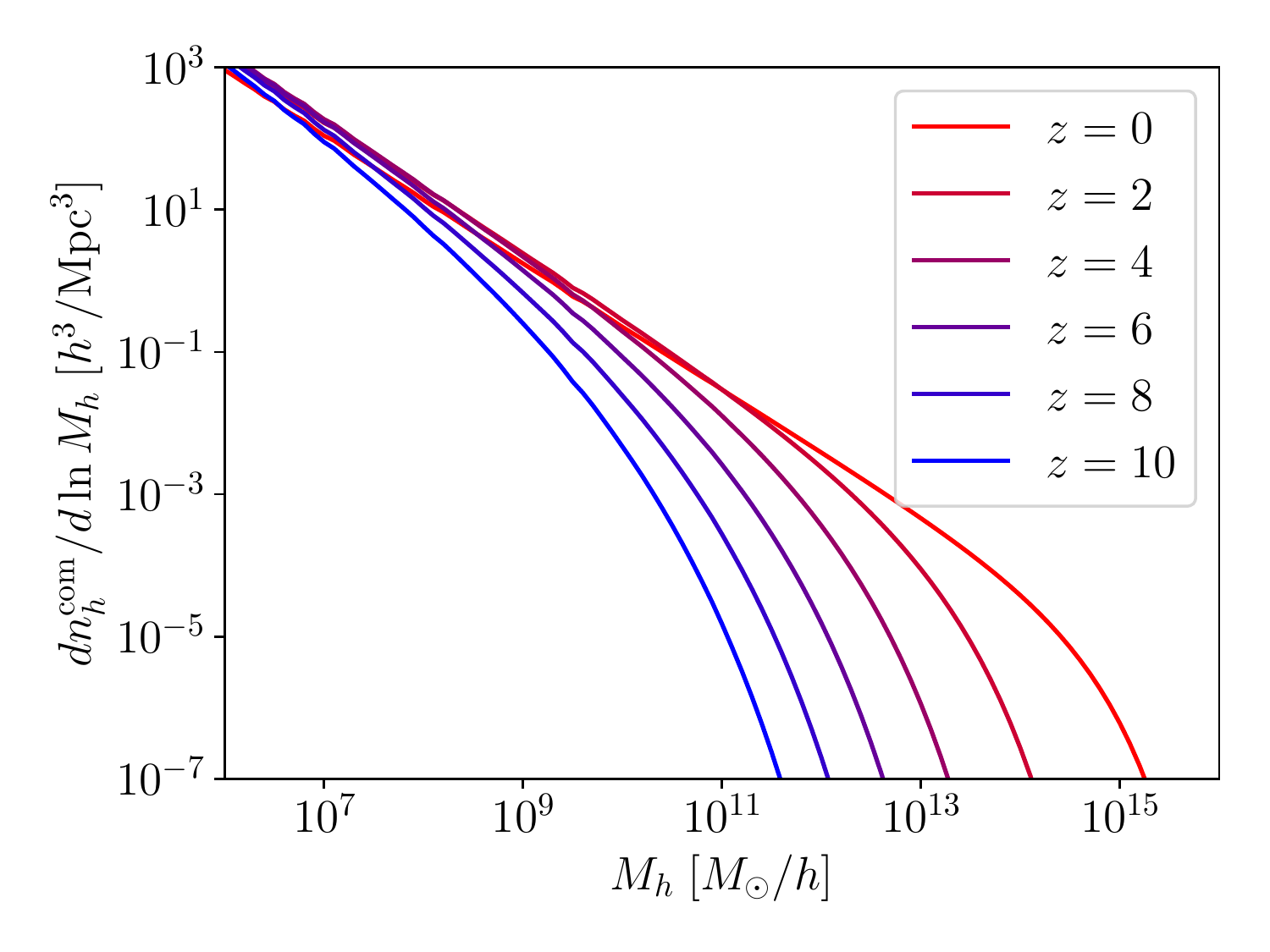}
    \caption{The Sheth-Tormen halo mass function in the comoving frame.}
    \label{fig:HMF}
\end{figure}

For the halo mass distribution, we adopt the Sheth-Tormen halo mass function \cite{Sheth:1999mn} which is an extension to the Press-Schechter formalism \cite{Press:1973iz} that fits well with the results of numerical simulations.
Examples of the halo mass function at several different redshifts are provided in FIG.~\ref{fig:HMF}.
Notice that the examples are shown in the comoving frame, rather than the physical volume.

\FloatBarrier

\subsection{Isotropic SGWB power spectral density from ABH-PBH merger}\label{sec:isotropic_result}

\begin{figure}[t]\centering
\includegraphics[width=0.5\textwidth]{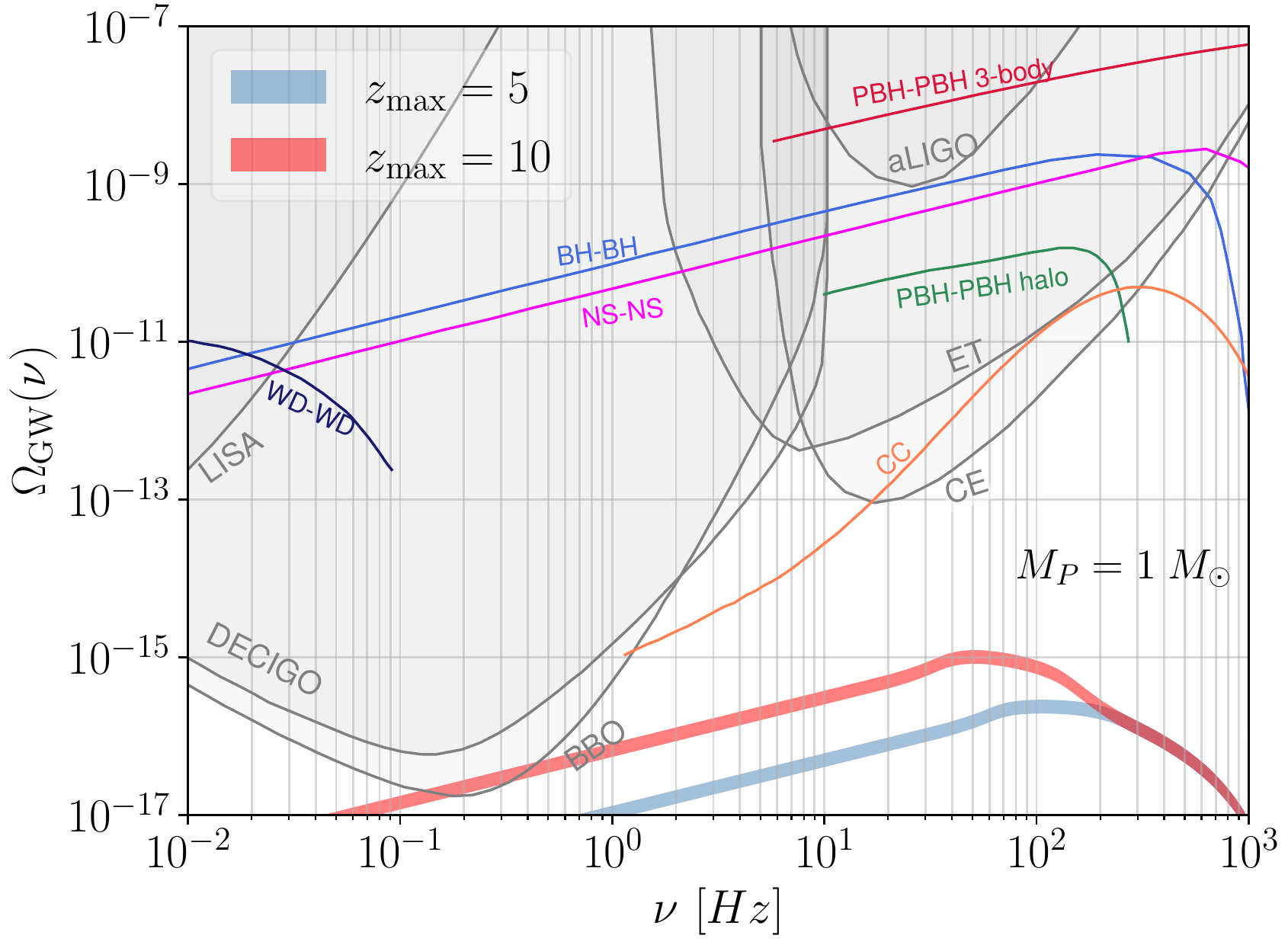}\includegraphics[width=0.5
\textwidth]{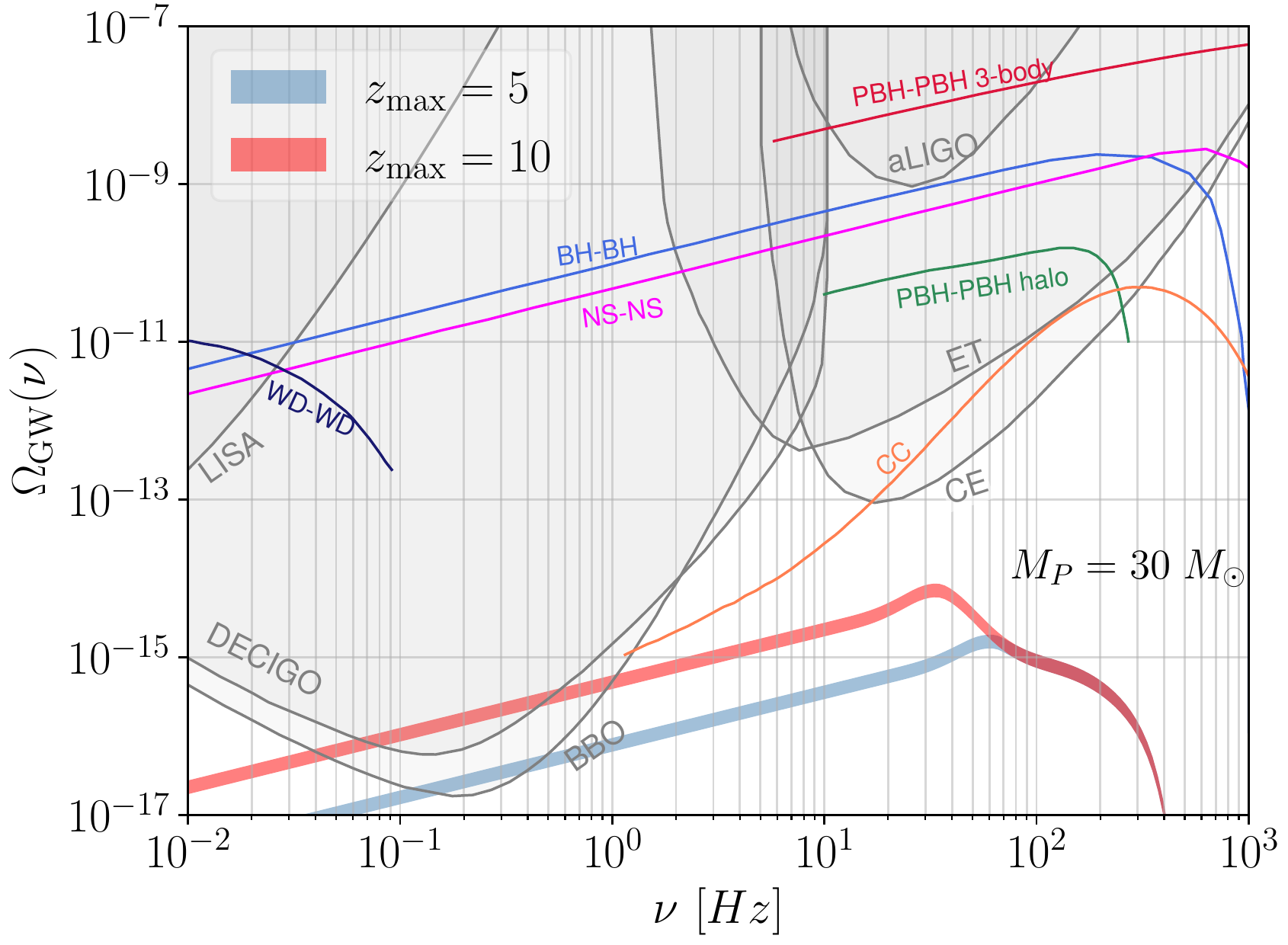}
\caption{Here we show the isotropic component of the stochastic GW background from ABH-PBH mergers. 
The left and right panels are results for primordial black hole mass as 1 and $30~M_{\odot}$, respectively.
For each colored band, the upper boundary is obtained by assuming $100\%$ elliptical galaxies, 
whereas the lower boundary assumes $100\%$ disk galaxies.
Various colored bands correspond to different choices of redshift cutoff, $z_{\rm max}$.
The sensitivities for several existing and future GW experiments \cite{Schmitz:2020syl} are shown as the gray curves.
We also present the expected SGWB produced by core collapse (CC) \cite{Crocker:2017agi} and other types of binaries, including PBH-PBH binaries formed at early time through 3-body process \cite{Wang:2016ana} (with chirp mass as $1~M_\odot,~f=0.05$), PBH-PBH binaries through gravitational capture in dark matter halos \cite{Mandic:2016lcn} (with chirp mass as $30~M_\odot,~f=1$), WD-WD binaries \cite{Farmer:2003pa}, as well as ABH-ABH and NS-NS binaries \cite{TheLIGOScientific:2016dpb}.
Note that we simply take representative results from these references and all the expected signals are subject to large uncertainties.
}\label{fig:spectrum_new}
\end{figure}

With all the ingredients prepared, the SGWB can be estimated by inserting Eq. (\ref{eq:gw_radiate_power}) and Eq.~(\ref{eq:rate}) into Eq.~(\ref{eq:master_formula_z}).
The SGWB energy density spectra are shown in Fig.~\ref{fig:spectrum_new}. 
Here the two benchmark values of the primordial black hole mass, $1~M_{\odot}$ and $30~M_{\odot}$, are presented by assuming $f=1$. 
Since the merger rate depends the PBH fraction $f$ linearly, results for different $f$ can be easily inferred.
Due to the large astrophysical uncertainties on the star population at high redshift, we present results with different choices of the cutoff redshift as $z_{\rm max}=5$ and $10$.

Notice that choosing different $z_{\rm max}$ has a noticeable effect.
Obviously, a larger $z_{\rm max}$ means more contribution from higher redshift which enhances the spectrum at lower frequency.
Consequently, we observe that the peak shifts to a lower frequency.
At a larger redshift, the validity of our astrophysical inputs may not be applicable.
Therefore, we do not extend our calculation to a redshift higher than $z=10$. 

Another noticeable difference for the two choices of $M_P$ is the overall magnitude of $\Omega_{\rm GW}$ --- the one with larger $M_P$ has a larger GW density spectrum.
This can be understood as follows.
The peak of $dE_{\rm GW}/d\ln \nu_s$ scales as $M_c^{5/3}\nu_{\rm merg}^{-1}\nu_{\rm ring}^{2/3}\sim M_A M_P$.
At the same time, the capture rate is proportional to $n_P\sigma_{\rm mer}$ which scales as $M_P^{-1} (M_A+M_P)^{10/7}M_A^{2/7}M_P^{2/7}$.
Clearly, a larger $M_P$ gives rise to a higher GW spectrum.

For each colored band in the plot, the upper boundary is obtained by assuming all elliptical galaxies, while the lower boundary corresponds to the assumption of $100\%$ disk galaxies.
The width of the band characterizes the uncertainty from the galaxy type.

The peak of the SGWB energy density from ABH-PBH mergers falls between \blue{$\mathcal{O}(10^{-15}-10^{-14})$}. This is far below the sensitivities of the existing or future ground-based experiments such as the aLIGO, Einstein Telescope (ET), Cosmic Explorer (CE).
For future space-based experiments, the Deci-Hertz Interferometer Gravitational-Wave Observatory (DECIGO) and the Big-Bang Observer (BBO) may have sensitivities to probe the ABH-PBH SGWB at a lower frequency band. However distinguishing such a spectrum from the astrophysical background remains challenging.

\section{PBH-ABH binaries in the Milky Way}\label{sec:MW}


The previous discussion is based on the isotropic distribution of galaxies in our Universe. The ABH-PBH mergers within our MW may also contribute to an anisotropic SGWB. These mergers are very close to us and we should be able to identify them one-by-one. However, if we focus on the frequency regime which is much lower than that of LIGO, it may still be useful to consider the SGWB produced during the inspiral stage. In this section, we estimate the binary formation rate in the MW and see whether it is proper to be treated as a source of SGWB.

\begin{table}\centering
 \begin{tabular}{c c c} 
 \hline\hline
 $\rho_{b,0}$ & $r_0$ & $r_{\rm cut}$\\ 
 $95.6~M_{\odot}~\rm pc^{-3}$ & $0.075~\rm kpc$ & $2.1~\rm kpc$\\ 
 \hline
 $\Sigma_{d,0,\rm thin}$ & $h_{d,\rm thin}$ & $R_{d,\rm thin}$\\
 $816.6~M_{\odot}~\rm pc^{-2}$ & $0.3~\rm kpc$ & $2.9~\rm kpc$\\
 \hline
 $\Sigma_{d,0,\rm thick}$ & $h_{d,\rm thick}$ & $R_{d,\rm thick}$\\
 $209.5~M_{\odot}~\rm pc^{-2}$ & $0.9~\rm kpc$ & $3.31~\rm kpc$\\
 \hline\hline
\end{tabular}
\caption{Parameters for MW components.}\label{tab:MW}
\end{table}

For $\rho_{\rm MW}$, we use the best-fitting mass model of the MW as well as its host halo provided in Ref.~\cite{McMillan:2011wd}.
In this model, the MW consists of three components --- the bulge, the thin disk and the thick disk.
The bulge and the disk density profiles take the following form respectively
\beqn
\rho_{b}&=&\frac{\rho_{b,0}}{(1+r'/r_0)^{\alpha}}~e^{-(r'/r_{\rm cut})^2}\,\,\,,\\
\rho_{d}&=&\frac{\Sigma_{d,0}}{2h_d}~e^{-\frac{\abs{h}}{h_d}-\frac{R}{R_d}}\,\,\,,
\eeqn
in which $\alpha=1.8$, $R$ and $h$ are the radius and the height in cylindrical coordinates, and
$r'=\sqrt{R^2+(h/q)^2}$
with the axis ratio $q=0.5$.
The dimensionful parameters are listed in Table \ref{tab:MW}.

For the host halo, we still take the NFW profile in Eq.~(\ref{eq:NFW}) with $\rho_0=0.00846~M_{\odot}/\rm pc^{3}$ and $R_s = 20.2~\rm kpc$.
We assume that the location of the solar system is right on the galactic disk ($h \simeq 0$) at a distance $R_{\odot}\simeq 8.29~\rm kpc$ away from the galactic center.

With these profiles, we can then estimate the binary formation rate in the MW.
Straightforward calculation shows that this rate is {$\sim 2.29 \times 10^{-12}~\rm yr^{-1}$ for $M_P=1~M_{\odot}$ and $\sim 8.32\times 10^{-13}~\rm yr^{-1}$ for $M_P=30~M_{\odot}$}.
The binary formation rate is so low that it is not likely to have even a single merger event during the age of the Universe.
Therefore, it is not appropriate to treat the ABH-PBH mergers in the MW as a source of SGWB.

\FloatBarrier

\section{Conclusion}\label{sec:conclusion}
In this paper, we study the SGWB produced by unresolved PBH-ABH mergers. We demonstrate that, in the higher frequency region, \ie $\mathcal{O}$(10-1000) Hz, the GW radiation is much lower than the reach of any existing or future ground-based GW experiment. In the lower frequency region, it may be within the reach of future space-based experiments such as DECIGO and BBO. Thus the SGWB produced by PBH-ABH mergers is not the key component leading to the discovery of PBHs. The uncertainty due to the type of galaxies (disk or elliptical) is relatively small. On the other hand, the uncertainty due to the choice of the cutoff redshift  has a noticeable effect in both the magnitude and the shape of the power spectrum.
In this paper, we assumed that all PBHs have the same mass.
In more realistic PBH models, the PBHs may have a broader mass spectrum.
Moreover, the duration between the bound state formation and the merger of the binary is neglected.
This is a safe approximation because it is much shorter than the time scale we are interested in \cite{Bird:2016dcv}.
{In addition, the PBHs are assumed to have a spatial distribution that follows the NFW profile.
A change in the spatial distribution, such as the clustering of PBHs, might help increase the merger rate \cite{DeLuca:2020jug}.}

The SGWB signal from PBH-ABH mergers are subject to large background.
For example, it is several orders of magnitude below the estimated backgrounds from WD-WD, NS-NS, and ABH-ABH mergers. In practice, it is very challenging to detect the SGWB signal from the PBH-ABH mergers.


Furthermore, the SGWB from PBH-ABH mergers also naturally constitute an inevitable background for PBH-PBH mergers.
Since the estimated PBH-ABH signal is much smaller than the PBH-PBH signal, our results illustrate that previous analysis for PBH-PBH mergers \cite{Mandic:2016lcn,Wang:2016ana} are still valid and not affected by this natural background.

\acknowledgments
We would like to thank Huanian Zhang and Zheng Zheng for useful discussions.
J.S. and F.H. are supported by the National Natural Science Foundation of China under Grants No. 12025507, 11690022, 11947302; and by the Strategic Priority Research Program and Key Research Program of Frontier Science of the Chinese Academy of Sciences under Grants No. XDB21010200, XDB23010000, ZDBS-LY-7003, and also by the CAS Project for Young Scientists in Basic Research under Grant No. YSBR-006.
F.H. is also supported by the National Science Foundation of China under Grants No. 12022514 and No. 11875003.  Y.Z. is
supported by U.S. Department of Energy under Award No. DESC0009959. Y.Z. would like to
thank the ITP-CAS for their kind hospitality.

\bibliographystyle{JHEP}
\bibliography{ref}

\end{document}